\documentclass[conference]{IEEEtran}
\IEEEoverridecommandlockouts
\usepackage{cite}
\usepackage{amsmath,amssymb,amsfonts}
\usepackage{algorithmic}
\usepackage{graphicx}
\usepackage{textcomp}
\usepackage{xcolor}
\usepackage{multirow}
\usepackage[center]{caption}
\usepackage{subcaption}
\usepackage{enumitem}

\def\BibTeX{{\rm B\kern-.05em{\sc i\kern-.025em b}\kern-.08em
    T\kern-.1667em\lower.7ex\hbox{E}\kern-.125emX}}
\begin{document}

\title{Citation Recommendations Considering Content and Structural Context Embedding \thanks{This work is partly supported by KAKEN (19H04116)}}

\author{\IEEEauthorblockN{1\textsuperscript{st} Yang Zhang}
\IEEEauthorblockA{\textit{Graduate School of Informatics} \\
\textit{Kyoto University}\\
Kyoto, Japan \\
zhang.yang.33z@st.kyoto-u.ac.jp}
\and
\IEEEauthorblockN{2\textsuperscript{nd} Qiang Ma}
\IEEEauthorblockA{\textit{Graduate School of Informatics} \\
\textit{Kyoto University}\\
Kyoto, Japan \\
qiang@i.kyoto-u.ac.jp}
}

\maketitle

\begin{abstract} 
The number of academic papers being published is increasing exponentially in recent years, and recommending adequate citations to assist researchers in writing papers is a non-trivial task. Conventional approaches may not be optimal, as the recommended papers may already be known to the users, or be solely relevant to the surrounding context but not other ideas discussed in the manuscript. In this work, we propose a novel embedding algorithm DocCit2Vec, along with the new concept of ``structural context'', to tackle the aforementioned issues. The proposed approach demonstrates superior performances to baseline models in extensive experiments designed to simulate practical usage scenarios.
\end{abstract}


\section{Introduction} \label{intro}
%[importance of citation recommendation]
When writing a paper, one of the common questions arising in many researchers' minds is that of which paper to cite for a certain context. However, with the exponentially increasing number of published papers, finding suitable citations is a considerable challenge.

\begin{figure}[tb]
    \centering
    \includegraphics[width=\columnwidth]{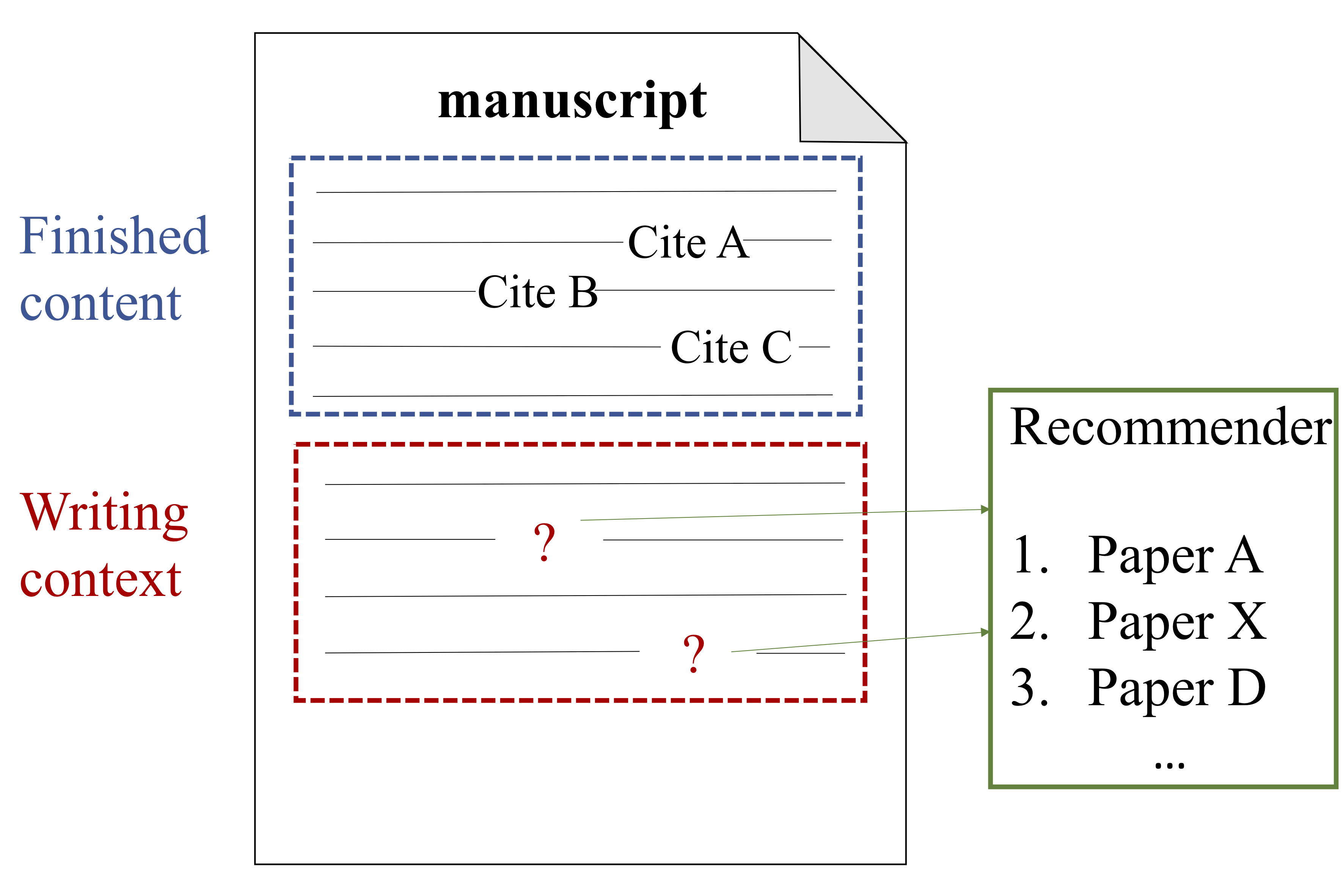}
    \caption{Citation recommendations for fleshing out content.} \label{fig0}
\end{figure}

%[different types of recommendation -> context recommendation]
Researchers generally rely on keyword-based recommenders, such as Google Scholar \footnote{https://scholar.google.com/} and Mendeley \footnote{https://www.mendeley.com/}. However, using keyword-based systems often results in unsatisfying results, as the query words may not carry adequate information to reflect the need to find relevant papers \cite{Jia:2017:ACR:3110025.3110150, 10.1007/978-3-319-76941-7_73}. Therefore, many approaches have been proposed to tackle this issue from various perspectives. For example, \cite{Caragea:2013:CSF:2467696.2467743,4061472,Jia:2017:ACR:3110025.3110150,Kucuktunc:2013:TPS:2492517.2492605,McNee:2002:RCR:587078.587096} attempt to recommend papers based on input seed papers. Some studies have considered personalized inputs, such as meta-data (title, abstract, keyword list, and publication year) \cite{Alzogbi2015PubRecRP,Li2018ConferencePR}. In addition, the studies \cite{P18-1222,He:2011:CRW:1935826.1935926,He:2010:CCR:1772690.1772734} proposed making recommendations based on the ``local context'' (the context surrounding a citation). Context-based recommendations are considered to be practical for aiding the paper-writing process.  However, due to the limitation of only utilizing the local context, these approaches may not be effective in some scenarios. For example, when a writer has finished writing a section of the manuscript, how can citations be effectively recommended to flesh out the content? Suppose that an author is writing a manuscript as in the example shown in Figure \ref{fig0}. Some parts of the manuscript are finished, and suitable citations have been inserted (content in the blue box). However, some parts have just been written, where citations have not yet been inserted (content in the red box). The objective of this work is to design a model to find relevant papers for freshly written content in a manuscript.

Conventional context-based approaches may not be optimal, for three reasons: 
\begin{enumerate}
    \item Previously cited papers are considered to be ``redundant'' for fleshing out the content. Users are already aware of these papers, and therefore they will cite them to flesh out the content without recommendation if they are relevant. 
    \item Previously cited papers contain information on co-citations, which may lead to more effective recommendations. For example, if the papers A, B, C, and D are frequently cited together, then it is likely that a paper will cite D if it has already cited A, B, and C. 
    \item An effective citation recommendation may not only be relevant to the idea discussed in the local context, but could also be linked to ideas that have been discussed in previous contexts.
\end{enumerate}

To provide a solution to the above-mentioned issues, we introduce the ``structural context'' in addition to the contexts containing word information (local contexts), which represents all the existing citations in the manuscript. This is motivated by the following considerations:
\begin{enumerate}
    \item The structural context carries information on all the existing citations, and can thus avoid recommending citations already known to the user.
    \item It can effectively recommend co-cited papers.
    \item Existing citations are relevant to various ideas in the fleshed out content.
\end{enumerate}

Considering these facts, we propose DocCit2Vec as an improved citation recommendation approach. This is a novel embedding algorithm, designed to embed academic papers as dense vectors by preserving information on the content, local context, and structural context.

The major contributions of this paper are summarized as follows: 
\begin{enumerate}
    \item The new concept of ``structural context'' is introduced to improve citation recommendations in practical scenarios (Section \ref{sec:pre}). 
    \item We propose the novel embedding model DocCit2Vec, and implement this with two neural network designs to explore the ability of an attention mechanism in citation recommendation tasks compared with a conventional average hidden layer (Section \ref{sec:model}). 
    \item The models are validated through extensive experiments, including three citation recommendation tasks and two classification tasks (Section \ref{sec:exp}). 
\end{enumerate}
The remainder of this paper is structured as follows. Section \ref{sec:re} summarizes related studies. Section \ref{sec:pre} provides notations, definitions, and the problem statement. Then, the embedding models and mathematical expressions are described in Section \ref{sec:model}, and the experiments and results are explained in Section \ref{sec:exp}. Finally, Section \ref{sec:sum} concludes the paper.

\section{Related Work} \label{sec:re}
\subsection{Document Embedding} \label{sec:emd}
Document embedding studies the representation of documents as continuous vectors. The Word2Vec model \cite{41224,Mikolov:2013:DRW:2999792.2999959} proposed a simple neural network to learn representation vectors for words in a given context. Doc2Vec \cite{pmlr-v32-le14} further expanded this model to embed documents. However, these two models suffer from information loss concerning links when embedding linked documents. HyperDoc2Vec \cite{P18-1222} was developed to incorporate the link information into dense vectors. Nevertheless, when applied to our scenario these techniques may still suffer from information loss, as they do not explicitly model the structural context.
%The underlying intuition is that similar words come with close representation vectors, for example that the result of {\itshape vector}(``{\itshape King}'') -  {\itshape vector}(``{\itshape Man}'') + {\itshape vector}(``{\itshape Woman}'') should be close to the {\itshape vector}(``{\itshape Queen}'')
\subsection{Citation Recommendation} \label{sec:cit}
Citation recommendation is the research field of finding the most relevant papers based on an input query. The studies \cite{Caragea:2013:CSF:2467696.2467743,4061472,Jia:2017:ACR:3110025.3110150,Kucuktunc:2013:TPS:2492517.2492605,McNee:2002:RCR:587078.587096} proposed utilizing a list of seed papers as a query, and providing recommendations through techniques such as collaborative filtering \cite{Caragea:2013:CSF:2467696.2467743,McNee:2002:RCR:587078.587096} and PageRank-based methods \cite{4061472,Jia:2017:ACR:3110025.3110150,Kucuktunc:2013:TPS:2492517.2492605}. Some studies have considered personalized queries, such as meta-data (title, abstract, keyword list, and publication year) \cite{Alzogbi2015PubRecRP,Li2018ConferencePR}. These methods would be effective in helping early-phase researches, such as by generating research ideas. However, for manuscript-writing tasks they may offer minimal help. Context-based recommendations \cite{P18-1222,He:2011:CRW:1935826.1935926,He:2010:CCR:1772690.1772734} take the context of a passage from a manuscript as a query, to provide recommendations aiming to assist the paper-writing process. However, these methods suffer from information loss, as information on citations already known to the user and other ideas in the paper are not taken into consideration. 

\subsection{Attention Mechanism} \label{sec:att}
The attention mechanism is commonly applied in vision-related tasks \cite{NIPS2014_5345}, where it can detect certain parts of an image to perform accurate predictions through learning. The authors of \cite{ling-etal-2015-contexts} extended the Word2Vec model using an attention mechanism to enhance the performance in classification tasks. In this work, we continue to explore the ability of the attention mechanism when applied in an embedding algorithm to enhance the performance for citation recommendation and classification tasks.

\section{Preliminaries} \label{sec:pre}
\subsection{Notations and Definitions} \label{sec:def}
% Let $\mathcal{X}$ be a collection of academic papers, which consists of a collection of document ids $\mathit{D}$ and a vocabulary $\mathit{W}$, \textit{i.e.}, $\mathcal{X} = \mathit{D} \cup \mathit{W}$. A single paper $\mathit{X} \in \mathcal{X}$ is associated with a document id $\mathit{d} \in \mathit{D}$, a collection of words $\mathit{\hat{W}} \subseteq \mathit{W}$, and cited documents (doc ids) $\mathit{\hat{D}} \subseteq \mathit{D}$, \textit{i.e.}, $\mathit{X} = \mathit{d} \cup \mathit{\hat{W}} \cup \mathit{\hat{D}}$.

We adapt the same notations to define hyper-documents as \cite{P18-1222}. Let $\mathit{w} \in \mathit{W}$ representing a word from a vocabulary $\mathit{W}$, and $\mathit{d}\in\mathit{D}$ representing a document id (the paper DOIs) from an id collection $\mathit{D}$. The textual information of a paper $\mathit{H}$ is represented as a sequence of words and document ids, \textit{i.e.}, $\mathit{\hat{W}} \cup \mathit{\hat{D}}$ where $\mathit{\hat{W}} \subseteq \mathit{W}$ and $ {\hat{D}} \subseteq \mathit{D}$. 

\begin{figure*}[tb]
  \centering
  \includegraphics[width=\textwidth]{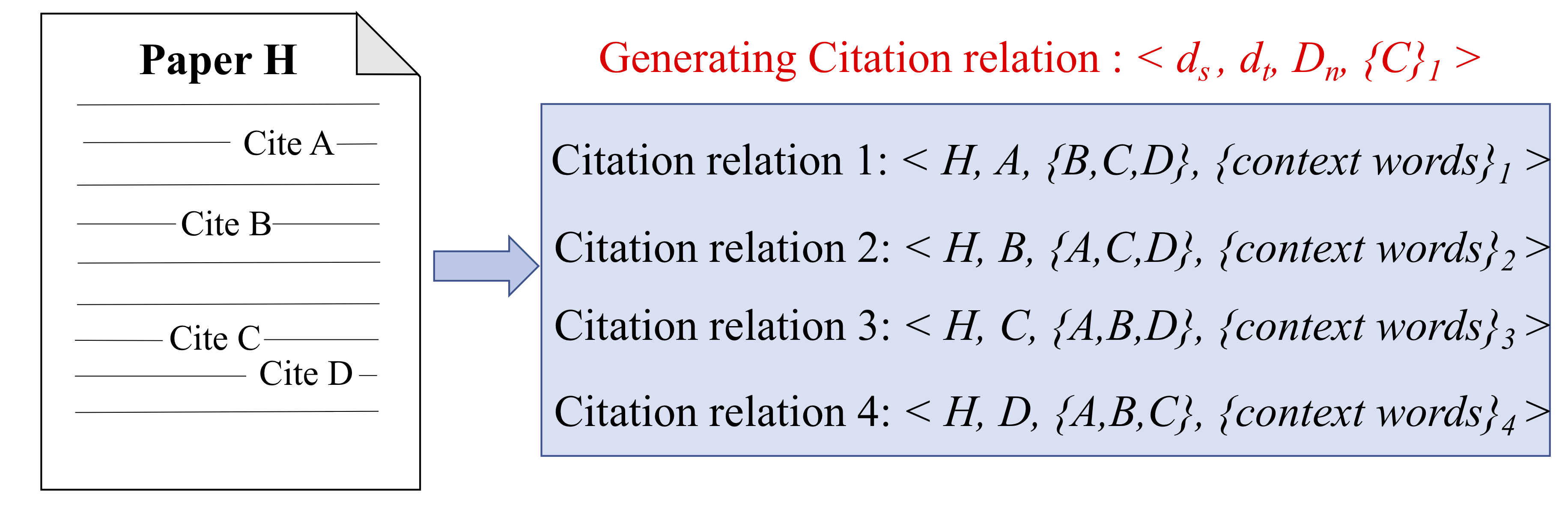}
  \caption{Citation relations of a paper \textit{H}.}
  \label{fig:citrelation}
\end{figure*}

Each citation relation (exemplified in Fig. \ref{fig:citrelation}) in a paper $\mathit{H}$ is expressed by a tuple $\langle \mathit{d_s}, \mathit{d_t}, \mathit{D_n}, \mathit{C}\rangle$, where $\mathit{d_s} \in \mathit{\hat{D}}$ is the id of the citing paper, the target id $d_t \in \mathit{\hat{D}}$ represents the cited paper, and $\mathit{C} \subseteq \mathit{\hat{W}}$ is the local context around $\mathit{d_t}$. If other citations exist in the manuscript, then these are defined as the \textit{``structural context''}, denoted by $\mathit{D_n}$ where $\{\mathit{d_n} | \mathit{d_n} \in \mathit{\hat{D}}, \mathit{d_n} \neq \mathit{d_t},  \mathit{d_n} \neq \mathit{d_s}\}$.

Embedding matrices are denoted as $\mathbf{D} \in \mathbb{R}^{k \times |D|}$ for documents and $\mathbf{W} \in \mathbb{R}^{k \times |W|}$ for words. The \textit{i}-th column of $\mathbf{D}$, $\mathbf{d_i}$, is a $\mathit{k}$-dimensional vector representing the document $\mathit{d_i}$, and the \textit{j}-th column of $\mathbf{W}$ is the \textit{k}-dimensional vector for the word $\mathit{w_j}$.

\subsection{Problem Statement}
There are two objectives for this task: First, embedding the papers as dense vectors, and second finding the most similar papers based on the embedded vector of an input paper. The two problems are stated as follows.

\subsubsection{Hyper-document Embedding}

Inspired by \cite{P18-1222}, we adopt two-step learning procedures to embed academic papers. The first step aims to embed the document and related contents, and the second embeds the local and structural contexts based on the learned vectors from the first step.

In the first step, the paper $\mathit{X}$ is simply treated as a plain text containing an id $\mathit{d_i}$ and words $\{w| w\in \mathit{\hat{W}} \}$. The objective is to learn to represent $\mathit{d_i}$ as a \textit{k}-dimensional vector $\mathbf{d_i}$, and all the words as a matrix $\mathbf{\hat{W}}$, where each vector $\mathbf{w_i}$ is the representation for a word $\mathit{w_i} \in \mathit{\hat{W}}$.

For the second step, a citation $\langle \mathit{d_s}, \mathit{d_t}, \mathit{D_n}, \mathit{C}\rangle$ from the paper $\mathit{X}$ and the previously learned matrices $\mathbf{\hat{W}}$ and $\mathbf{d_i}$ are given. Then, each document id in the set $\mathit{d_i} \in \mathit{d_s} \cup \mathit{d_t} \cup \mathit{D_n}$ is learned, and these are represented by the matrix $\mathbf{\hat{D_c}}$, where each column is the \textit{k}-dimensional vector for the corresponding doc id. Furthermore, $\mathbf{\hat{W_c} \subseteq \mathbf{\hat{W}}}$ represents the embedding matrix for the context words $\mathit{C}$, where each column is a vector representing  $\mathit{w} \in \mathit{C}$.

\subsubsection{Citation Recommendation} \label{cases}

We set three usage cases to simulate the proposed scenario. For each case, the purpose is to find the most similar papers based on vector similarities.

\begin{itemize}
    \item Case 1: In this case, it is assumed that the author has inserted a number of citations in the completed content of the manuscript. To find the most similar paper, we take the local context of the content without citations and the structural context (previously existing citations) as input vectors, and rank document vectors by similarity.
    \item Case 2: This case assumes that some of the existing citations are invalid, because they might not be available in the dataset or the author has made typos. When recommending, the local context and randomly selected structural contexts are adopted for the similarity computation.
    \item Case 3: If all the existing citations are invalid or the author has not inserted any citations, then only the local context is adopted for the similarity computation.
\end{itemize}

%%It is assumed that the paper $\mathit{X}$ comes with a set of necessary citations $\mathit{N}\subseteq\mathit{D}$, whereas $\mathit{U}$ denotes a set of unnecessary citations where $\mathit{U}\subseteq\mathit{D}$, and $\mathit{U}\cup\mathit{N} = \emptyset$.

\section{DocCit2Vec} \label{sec:model}

\subsection{Representing Documents and Citations}

The embedding of documents refers to the process of assigning vectors to each document id and word, and optimizing these with predictions of target words or document ids. For example, \textit{pv-dm} \cite{pmlr-v32-le14} learns two vectors (IN and OUT vectors) for each word, \textit{i.e.}, $\mathbf{w^I}$ and $\mathbf{w^O}$, respectively, and one OUT vector $\mathbf{d^I}$ for each document id. Given a document id and its content, the model picks a word as the target, and averages over the IN vectors of the document id $\mathbf{d^I}$ and all the words in the surrounding context $\{\mathbf{w^I_i} | i \in \mathit{C}\}$ to make a prediction of the target word's OUT vector $\mathbf{w^O}$. Representing documents as predictions of target words enables the document vector to reflect the words the document contains, \textit{i.e.}, the content.

DocCit2Vec conceptualizes the learning process as the prediction of a target citation, so that an embedded document vector carries the information of a target citation. This process is illustrated in Fig.\ref{model}. Given a citation relation, DocCit2Vec picks one publication from the citation list as the target, and utilizes the surrounding context and structural context as known knowledge to maximize the occurrence of the target citation by updating the parameters (\textit{i.e., the embedding vectors}) of the neural network. The model learns two document embedding vectors IN and OUT, where the IN vector $\mathbf{d^I}$ characterizes the document as a citing paper and the OUT vector $\mathbf{d^O}$ encodes its role as a cited paper \cite{P18-1222}. In addition, the model learns IN word vectors $\mathbf{w^I}$.

As mentioned in Section 3.2, embedding academic papers involves two steps: embedding of the content and of the citations. For the first step, we adopt the retrofitting technique as in \cite{P18-1222}, which initializes a predefined number of iterations based on the \textit{pv-dm} model and then utilizes the learned vectors as the ``base'' vectors for step two.

Two designs of DocCit2Vec are proposed, the first design (DocCit2Vec-avg) uses an conventional averaged hidden layer, and the second (DocCit2Vec-att) adopts an attention hidden layer.

\subsection{DocCit2Vec with an Average Hidden Layer}
\begin{figure*}[tb]
    \centering
    \includegraphics[width=\textwidth]{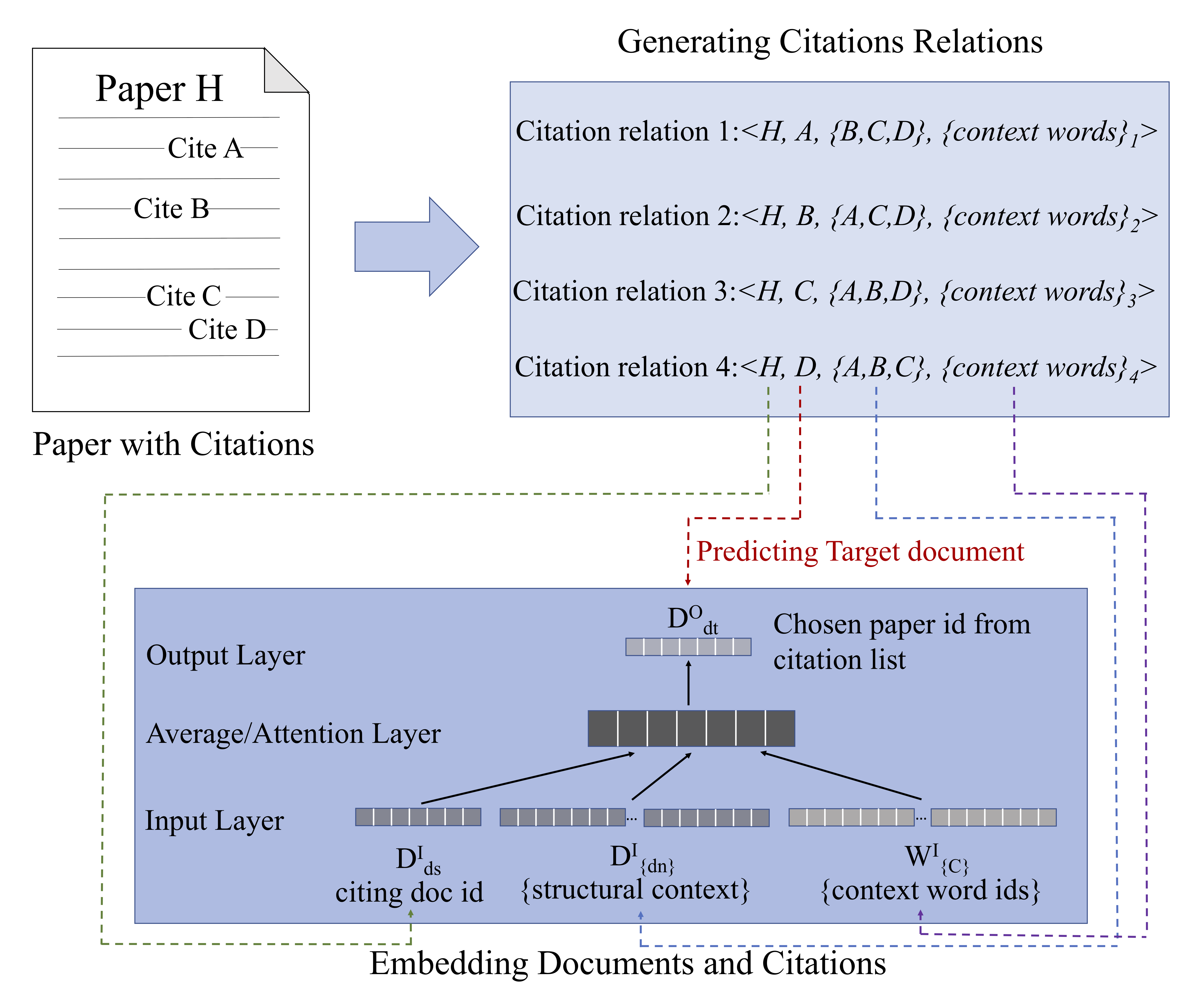}
    \caption{Overview of DocCit2Vec} \label{model}
\end{figure*}

The architecture of DocCit2Vec-avg is constructed based on the \textit{pv-dm} structure of Doc2Vec \cite{pmlr-v32-le14} and HyperDoc2Vec \cite{P18-1222}, shown on the right side of Fig. \ref{model}. The model initializes an IN document matrix $\mathbf{D^{I}}$ at the input layer, an OUT document matrix $\mathbf{D^{O}}$ at the output layer, and an IN word matrix $\mathbf{W^{I}}$ at the input layer. To obtain a citation relation $\langle \mathit{d_{x}},\mathit{d_t}, \{\mathit{d_i}|i\in \mathit{D_n}\}, \{w|w\in \mathit{C}\}\rangle$, the model averages over the corresponding IN vectors of $\mathit{d_{x}}$, $\{\mathit{d_i}|i\in \mathit{D_n}\}$, and $\{w|w\in \mathit{C}\}$. The output layer is computed using a multi-class softmax classifier, and the output value is regarded as the probability of the occurrence of $\mathit{d_t}$.

To learn all the citation relations $\mathcal{C}$, the model is statistically expressed as
\begin{equation}
    \max_{\mathbf{D^{I}}, \mathbf{D^{O}}, \mathbf{W^{I}}} 
    \frac{1}{|\mathcal{C}|} \sum_{\langle \mathit{d_s},\mathit{d_t}, \mathit{D_n}, \mathit{C}\rangle \in \mathcal{C}}
    \log P(\mathit{d_t|\mathit{d_s}, \mathit{D_n}, \mathit{C}})
\end{equation}

The hidden layer of the neural network is expressed as
\begin{equation}
    \mathbf{x} = \frac{1}{1 + |\mathit{D_n}| + |\mathcal{C}|}
    \Big(\mathbf{d_s}^{I} + \sum_{\mathit{d_n} \in \mathit{D_n}}\mathbf{d_n}^{I} + \sum_{\mathit{w} \in \mathit{C}} \mathbf{w}^{I}\Big)
\end{equation}

The output layer adopts a multi-class softmax function, which is represented as
\begin{equation}
    P(\mathit{d_t}|\mathit{d_s}, \mathit{D_n}, \mathit{C}) = 
    \frac{\exp \big(\mathbf{x}^T \mathbf{d}^O_t \big)}{\sum_{\mathit{d}\in\mathit{D}} \exp \big(\mathbf{x}^T\mathbf{d}^O \big)}
\end{equation}

The negative sampling technique \cite{Mikolov:2013:DRW:2999792.2999959} is adopted to optimize the efficiency of the training procedure:
\begin{equation}
    \log \sigma \big( \mathbf{x}^T\mathbf{d}^O_t \big) +
    \sum^n_{i=1} \mathcal{E}_{\mathit{d_i} ~ P_{N(d)}} \log \sigma \big(-\mathbf{x}^T\mathbf{d}^O_i \big)
\end{equation}

To optimize the model, the gradients of the loss function with respect to the parameters $\mathbf{D}^I$, $\mathbf{D}^O$, and $\mathbf{W}^I$ are computed. The parameters are then updated with an input learning rate through backpropagation.

\subsection{DocCit2Vec with an Attention Hidden Layer}
The architecture of DocCit2Vec-att is the same as that of DocCit2Vec-avg on the right side of Fig. \ref{model}, except that the averaged hidden layer is replaced by an attention layer, inspired by \cite{ling-etal-2015-contexts}. In addition to the original parameters, the weight vector $\mathbf{K} \in \mathbb{R}^{1 \times (|D|+|W|)}$ is introduced at the attention layer, where each value denotes the importance of a word or document. The model is statistically expressed as

\begin{equation}
    \max_{\mathbf{D^{I}}, \mathbf{D^{O}}, \mathbf{W^{I}}, \mathbf{K}} 
    \frac{1}{|\mathcal{C}|} \sum_{\langle \mathit{d_s},\mathit{d_t}, \mathit{D_n}, \mathit{C}\rangle \in \mathcal{C}}
    \log P(\mathit{d_t|\mathit{d_s}, \mathit{D_n}, \mathit{C}})
\end{equation}

Instead of the averaged hidden layer, the attention layer computes a weighted sum of an individual word and document through the multiplication of the vector and its weight ratio, which is expressed as follows:
\begin{equation}
    \mathbf{x} = 
    \mathbf{d_s}^{I} \cdot \mathit{a_{d_s}}+ \sum_{\mathit{d_n} \in \mathit{D_n}}\mathbf{d_n}^{I} \cdot \mathit{a_{d_n}} + \sum_{\mathit{w} \in \mathit{C}} \mathbf{w}^{I} \cdot \mathit{a_{w}}
\end{equation}

The terms $\mathit{a_{d_s}}$, $\mathit{a_{d_n}}$, and $\mathit{a_{w}}$ are associated weight ratios for the documents $\mathit{d_s}$ and $\mathit{d_n}$ and the word $\mathit{w}$. The weight ratios are computed by using the matrix $\mathbf{K}$ as follows:

\begin{equation}
    \mathit{a_i} = \frac{\exp k_i}{\sum_{j \in (\mathit{d_s} \cup \mathit{D_n} \cup \mathit{C})} \exp k_j}
\end{equation}

For output layer, negative sampling is performed in an identical manner to DocCit2Vec. In addition, the gradient of $\mathbf{K}$ is computed to optimize DocCit2Vec using attention.

\section{Experiments} \label{sec:exp}
\subsection{Datasets and Experimental Settings}
Two pre-processed datasets DBLP and ACL Anthology \cite{P18-1222}, containing full texts of academic papers in the computer science domain are utilized. Statistical summaries of the datasets are provided in Table \ref{tab:data}.
%\vspace{-3em}
\begin{table*}[tb]
    \centering
    \caption{Statistics of the Datasets} \label{tab:data}
    \begin{tabular}{l|l|ll|l|ll|l}
    \hline\hline
    \textbf{Dataset}      & \multicolumn{3}{l|}{\textbf{No. of Docs}}   & \multicolumn{3}{l|}{\textbf{No. of Citations}}  & \textbf{Experiments}                                                                                    \\ \hline
    \multirow{2}{*}{ACL}  & \multirow{2}{*}{20,408}  & Train: & 14,654  & \multirow{2}{*}{108,729}   & Train: & 79,932    & \multirow{2}{*}{Citation Rec.}                                                                \\ \cline{3-4} \cline{6-7}
                          &                          & Test:  & 1,563   &                            & Test:  & 28,797    &                                                                                                         \\ \hline
    \multirow{2}{*}{DBLP} & \multirow{2}{*}{649,114} & Train: & 630,909 & \multirow{2}{*}{2,874,303} & Train: & 2,770,712 & \multirow{2}{*}{\begin{tabular}[c]{@{}l@{}}Citation Rec.\\ Classifications\end{tabular}} \\ \cline{3-4} \cline{6-7}
                          &                          & Test:  & 18,205  &                            & Test:  & 103,591   &                                                                                                         \\ \hline\hline
    \end{tabular}
\end{table*}
%\vspace{-2em}
We conduct two experiments on the large-sized DBLP dataset: citation recommendations and topic classification. For the citation recommendations, we picked all the documents with more than one citation from the same dataset published in recent years as the test dataset, and the remainder of the dataset was utilized to train the embedding model. Another medium-sized dataset, the ACL Anthology, was also adopted to implement the citation recommendation experiment. This contains a similar number of average citations per document as DBLP. Same to DBLP, we selected a test dataset for testing the recommendation performance, and a training dataset for to train the model.The titles and included citations of the texts are replaced by indices, so that the algorithm can detect these as hyperlinks. For example, a citation [1] or (Song et al. 2018) is replaced with an index, and the title of the cited paper is replaced by the same index. Following the same settings as in \cite{P18-1222}, we set 50 words before and after a citation as the citation context.

We implemented three baseline models: Word2Vec, Doc2Vec, and HyperDoc2Vec. The Gensim package \cite{rehurek_lrec} was utilized to implement the baseline models, as well as the foundation for developing DocCit2Vec. We employed the same hyper-parameter settings as in \cite{P18-1222}. For Word2Vec, the embedding size was set to 100 with a window size of 50 and 5 epochs, using the \textit{cbow} structure, and the default Gensim settings were followed. For Doc2Vec, the same embedding, window size, and epoch setting were adopted with the \textit{pv-dbow} structure. For HyperDoc2Vec, the same embedding and window size were adopted, with 100 iterations and 1000 negative samplings, and with the initialization of Doc2Vec at five epochs. The same settings were adopted for DocCit2Vec as for HyperDoc2Vec. The model are implemented on a Linux server with 12 cores of Intel Xeon E5-1650 cpu and 128Gb memory installed with Anaconda 5.2.0 and Gensim 2.3.0.

\subsection{Recommendation Experiments}
As mention in Section \ref{cases}, we designed three cases for the citation recommendation experiment. Each case adopts different input query:
\begin{itemize}
    \item Case 1: Utilize the averaged vector of the context words (50 words before and after a citation) and structural contexts. 
    \item Case 2: Utilize the averaged vector of the context words and randomly selected structural contexts.
    \item Case 3: Utilize the averaged vector of the context words. 
\end{itemize}
We adopt different similarity calculation methods for the best performance of each model, as described in \cite{P18-1222}. We employ the IN-for-OUT (I4O) method for Word2Vec (W2V) and HyperDoc2Vec (HD2V), which uses the averaged IN feature vectors to rank the OUT document vectors by the dot product. Doc2Vec implements an IN-for-IN (I4I) method, which first infers a vector from the IN feature vectors by applying the learned model, and then ranks the IN document vectors by the cosine similarity. In addition, as in \cite{P18-1222}, we run Doc2Vec-nc (D2V-nc) on the training file without citations, and Doc2Vec-cacNev (D2V-cac) on the training file with ``augmented contexts'', \textit{i.e.}, each citation context is copied into the target document. For DocCit2Vec-avg (DC2V-avg) and DocCit2Vec-att (DC2V-att), we employ the same I4O method. For the evaluation, the recall, MAP, and nDCG are reported and compared for the top 10 results.

\begin{table*}[tb]
    \centering
    \caption{Citation Recommendation Results} 
    \label{tab:result1}
    \begin{tabular}{l|lll|lll}
    \hline\hline
    \multirow{2}{*}{\textbf{Model}} & \multicolumn{3}{l|}{\textbf{ACL Anthology}}      & \multicolumn{3}{l}{\textbf{DBLP}}                                    \\ \cline{2-7} 
                                    & Recall         & MAP            & nDCG           & Recall         & MAP            & nDCG                               \\ \hline
     
    W2V (Case 1)                & 27.25          & 13.74          & 19.51          & 20.47          & 10.54          & 14.71                              \\
    W2V (Case 2)                & 26.50          & 13.74          & 19.51          & 20.47          & 10.55          & 14.71                              \\  
    W2V (Case 3)                & 26.06          & 13.21          & 18.66          & 20.15          & 10.40          & 14.49                              \\
    D2V-nc (Case 1)             & 19.92          & 9.06           & 13.39          & 7.90           & 3.17           & 4.96                               \\
    D2V-nc (Case 2)             & 19.89          & 9.06           & 13.38          & 7.90           & 3.17           & 4.96                               \\
    D2V-nc (Case 3)             & 19.89          & 9.07           & 13.38          & 7.91           & 3.17           & 4.97                               \\
    D2V-cac (Case 1)            & 20.51          & 9.24           & 13.68          & 7.91           & 3.17           & 4.97                               \\
    D2V-cac (Case 2)            & 19.89          & 9.06           & 13.38          & 7.91           & 3.17           & 4.97                               \\
    D2V-cac (Case 3)            & 20.51          & 9.24           & 13.69          & 7.89           & 3.17           & 4.97                               \\
    HD2V (Case 1)               & \textbf{37.53} & 19.64          & 27.20          & 28.41          & 14.20          & 20.37                              \\
    HD2V (Case 2)               & \textbf{36.85} & \textbf{19.64} & \textbf{27.20} & 28.43          & 14.20          & 20.39                              \\ 
    HD2V (Case 3)               & \textbf{36.24} & \textbf{19.32} & \textbf{26.79} & 28.41          & 14.20          & 20.37                              \\ \hline
    DC2V-att (Case 1)           & 27.48          & 13.42          & 19.24          & 7.381          & 3.358          & 4.889                              \\
    DC2V-att (Case 2)           & 25.01          & 12.37          & 17.66          & 6.06           & 2.73           & 3.98                               \\
    DC2V-att (Case 3)           & 25.01          & 11.48          & 16.27          & 5.20           & 2.36           & 3.43                               \\
     DC2V-avg (Case 1)           & 36.89          & \textbf{20.44} & \textbf{27.72} & \textbf{44.23} & \textbf{21.80} & \textbf{31.34}                     \\
     DC2V-avg (Case 2)           & 33.67          & 18.40          & 25.10          & \textbf{40.37} & \textbf{20.15} & \textbf{28.69}    \\
    DC2V-avg (Case 3)           & 31.14          & 16.97          & 23.20          & \textbf{40.37} & \textbf{19.02} &  \multicolumn{1}{r}{\textbf{26.84}} \\ \hline\hline
    \end{tabular}
    \vspace{-4mm}
\end{table*}
Four observations can be made from the results (Table \ref{tab:result1}). First, DocCit2Vec-avg demonstrated a superior performance on the larger sized dataset, DBLP, with a significant improvement. The recall was higher by approximately 12\% to 15\% according to the different cases compared to the second-best model HyperDoc2Vec, with 4\% to 7\% improvements for the MAP and 6\% to 11\% for the nDCG. Second, all the models yielded better performances for the medium-sized dataset ACL Anthology, except for DocCit2Vec-avg, which may reveal that DocCit2Vec-avg requires a larger volume of data to converge. HyperDoc2Vec yielded the best results for this dataset, followed by DocCit2Vec-avg with close scores. Third, all baseline models exhibit similar scores across the three cases except for DocCit2Vec-avg. It is observed that DocCit2Vec-avg constantly yielded the best scores for the first case, where all the structural contexts are included, and the second best for the second case, where the structural contexts are randomly picked. This indicates that the information on the structural contexts is embedded into the embedding vectors. Fourth, the performances of DocCit2Vec-att are among the lowest, suggesting that citation recommendation is not a suitable task for this model.

\subsection{Classification Experiments}
We conduct two classification experiments: topic classification \footnote{Cora dataset, https://people.cs.umass.edu/\~{}mccallum/data.html.} and classification of the functionality of citations \footnote{Dataset from \cite{Teufel:2006:ACC:1610075.1610091}.}. The two experiments aim to test the performances of document and word vectors separately, where the first experiment adopts document vectors and the second uses words vectors. The dataset for topic classification includes 5,975 academic papers and 10 unique fields to which they belong. The dataset for the classification of the functionality of citations includes 2,824 citation contexts, each with a classified functionality, such as ``PBas: Cited work used as a basis or starting point'' , which represent 12 unique classes.

We employ three methods for the first experiment: The first uses a concatenation of IN and OUT vectors of the documents (``IN+OUT'' in Table \ref{tab:result2}), the second uses the IN vectors of the documents, and the third concatenates the IN or IN+OUT vectors with embedding vectors from DeepWalk \cite{Perozzi:2014:DOL:2623330.2623732}. For the second experiment, we utilize the averaged IN vector of the context words.

\begin{table*}[tb]
    \centering
    \caption{Results of Classification Experiments} \label{tab:result2}
    \begin{tabular}{l|ll|ll|ll}
    \hline\hline
    \multirow{3}{*}{\textbf{Model}} & \multicolumn{4}{l|}{\textbf{Topic Classification}}                                              & \multicolumn{2}{l}{\textbf{\begin{tabular}[c]{@{}l@{}}Classification of \\ Functionality\end{tabular}}} \\ \cline{2-7} 
                                    & \multicolumn{2}{l|}{Original}                  & \multicolumn{2}{l|}{with DeepWalk}             & \multicolumn{1}{l|}{\multirow{2}{*}{F1-micro}}                      & \multirow{2}{*}{F1-macro}                     \\ \cline{2-5}
                                    & \multicolumn{1}{l|}{F1-micro} & F1-macro       & \multicolumn{1}{l|}{F1-micro} & F1-macro       & \multicolumn{1}{l|}{}                                               &                                               \\ \hline
    W2V (IN+OUT)                    & 58.11                         & 37.00          & 74.62                         & 63.66          & N/A                                                                 & N/A                                           \\
    W2V (IN)                        & 57.83                         & 36.90          & 76.61                         & 66.94          & \textbf{77.36}                                                      & \textbf{56.52}                                \\
    D2V-nc                          & 78.70                         & 70.91          & 82.67                         & 76.16          & 63.51                                                               & 6.47                                          \\
    D2V-cac                         & 78.99                         & 71.11          & 82.41                         & 75.95          & 63.51                                                               & 6.47                                          \\
    HD2V (IN+OUT)                   & 80.26                         & 73.72          & 82.38                         & 76.11          & N/A                                                                 & N/A                                           \\
    HD2V (IN)                       & 79.12                         & 72.78          & 82.38                         & 76.30          & 75.63                                                               & 54.33                                         \\
    DC2V-att (IN+OUT)               & 82.70                         & 76.86          & 84.84                         & 79.50          & N/A                                                                 & N/A                                           \\
    DC2V-att (IN)                   & \textbf{83.80}                & \textbf{78.43} & \textbf{85.49}                & \textbf{80.59} & 74.46                                                               & 54.43                                         \\
    DC2V-avg (IN+OUT)               & 77.56                         & 70.43          & 79.23                         & 72.65          & N/A                                                                 & N/A                                           \\
    DC2V-avg (IN)                   & 75.28                         & 68.18          & 77.28                         & 70.36          & 75.91                                                               & 54.67   \\ \hline\hline                           
    \end{tabular}
\end{table*}

DocCit2Vec-att yields the best performance in topic classification (Table \ref{tab:result2}), with F1-macro and F1-micro scores approximately 3\% to 4\% higher compared to the second-best model HyperDoc2Vec. DocCit2Vec-avg is ranked as the second lowest among all the models. For the second experiment, neither DocCit2Vec-att nor DocCit2Vec-avg is ranked at the top. First, the results reveal that DocCit2Ve-att improves the classification abilities of document embedding vectors, although not word embedding vectors. Second, the reason for the inferiority of DocCit2Vec-avg is that it takes multiple documents as input, and therefore it emphasizes the ``similarity'' between documents. In summary, attention focuses on ``difference'', whereas the averaged layer emphasizes ``similarity''.

\section{Conclusion} \label{sec:sum}
We propose a novel document embedding model, DocCit2Vec, with consideration of the ``structural context'', to improve the recommendation performance for writing academic papers. Two implementations of the model are proposed: the first one comes with an average hidden layer (DocCit2Vec-avg) and the second with an attention hidden layer (DocCit2Vec-att). Experimental results demonstrate the superior performance of DocCit2Vec-avg for citation recommendation tasks. Furthermore, DocCit2Vec-att yields an effective performance for the classification task with the adoption of document embedding 
vectors. In the future, we plan to design more sophisticated neural networks, and combine with graph and keyword based approaches to further improve the performance.

\bibliographystyle{IEEEtran}
\bibliography{citations_new}

% \begin{thebibliography}{00}
% \bibitem{b1} G. Eason, B. Noble, and I. N. Sneddon, ``On certain integrals of Lipschitz-Hankel type involving products of Bessel functions,'' Phil. Trans. Roy. Soc. London, vol. A247, pp. 529--551, April 1955.
% \bibitem{b2} J. Clerk Maxwell, A Treatise on Electricity and Magnetism, 3rd ed., vol. 2. Oxford: Clarendon, 1892, pp.68--73.
% \bibitem{b3} I. S. Jacobs and C. P. Bean, ``Fine particles, thin films and exchange anisotropy,'' in Magnetism, vol. III, G. T. Rado and H. Suhl, Eds. New York: Academic, 1963, pp. 271--350.
% \bibitem{b4} K. Elissa, ``Title of paper if known,'' unpublished.
% \bibitem{b5} R. Nicole, ``Title of paper with only first word capitalized,'' J. Name Stand. Abbrev., in press.
% \bibitem{b6} Y. Yorozu, M. Hirano, K. Oka, and Y. Tagawa, ``Electron spectroscopy studies on magneto-optical media and plastic substrate interface,'' IEEE Transl. J. Magn. Japan, vol. 2, pp. 740--741, August 1987 [Digests 9th Annual Conf. Magnetics Japan, p. 301, 1982].
% \bibitem{b7} M. Young, The Technical Writer's Handbook. Mill Valley, CA: University Science, 1989.
% \end{thebibliography}
%\vspace{12pt}

\end{document}